\documentclass[a4paper,12pt]{article}

\usepackage{amsmath,amssymb}
\usepackage[dvips]{graphicx}

\newcommand{\del}{\partial}
\newcommand{\vtilde}{\tilde{v}}
\newcommand{\nbar}{\bar{n}}
\newcommand{\alphat}{\tilde{\alpha}}
\newcommand{\Xdot}{\dot{X}}
\newcommand{\Xcal}{\mathcal{X}}
\newcommand{\Teff}{T_{\mathrm{eff}}}
\newcommand{\Jcal}{\mathcal{J}}
\newcommand{\Acal}{\mathcal{A}}
\newcommand{\Bcal}{\mathcal{B}}

\begin{document}
\renewcommand{\thefootnote}{\fnsymbol{footnote}}
\begin{titlepage}
\hfill
{\hfill \begin{flushright} 
HIP-2011-25/TH
\end{flushright}
}

\vspace*{10mm}

\begin{center}
{\LARGE \textbf{
Transport coefficients of D1-D5-P system and the membrane paradigm}}
\vspace*{18mm}

\large{Yuya Sasai} \footnote{E-mail: sasai@mappi.helsinki.fi}

\vspace*{13mm}
\large{\textit{
Helsinki Institute of Physics and Department of Physics \\
University of Helsinki,  
P.O.Box 64, FIN-00014 Helsinki, Finland \\
}}

\end{center}

\vspace*{20mm}

\begin{abstract}
I discuss a correspondence between string theory and the black hole membrane paradigm  in the context of the D1-D5-P system. By using  the Kubo formula, I calculate  transport coefficients of the effective string model induced by two kinds of minimal scalars. Then, I show that these transport coefficients \textit{exactly} agree with the  corresponding membrane transport coefficients of a five-dimensional near-extremal black hole with three charges.

\end{abstract}

\end{titlepage}

\newpage
\renewcommand{\thefootnote}{\arabic{footnote}}
\setcounter{footnote}{0}

\section{Introduction}
In recent decades, much progress has been made on a correspondence between  a black hole and string theory. In \cite{Susskind:1993ws,Sen:1995in}, the Bekenstein-Hawking entropy of a  black hole has been derived from a highly excited fundamental string up to a numerical factor. These works have been generalized and  it has been conjectured that
 a highly excited fundamental string  becomes a black hole with the same mass and charges when the string coupling  is  increased and becomes a critical value which is called the correspondence point \cite{Horowitz:1996nw}.

Although the correct numerical factor of the black hole entropy could not be reproduced from the fundamental string,  the Bekenstein-Hawking entropy of a five-dimensional extremal black hole with three charges has been found to be  exactly equal to  the degeneracy of BPS  states in a system which is composed of $n_1$ D1-branes wrapped on $S^1$   and $n_5$ D5-branes wrapped on $S^1\times M_4$, where $M_4=K3$ or $T^4$ \cite{Strominger:1996sh,Callan:1996dv}. This  system is called the D1-D5-P system. In the case of $M_4=T^4$, the microscopic states are effectively described by 
  a single D1-brane with winding number $n_1n_5$ which vibrates only inside $T^4$. It has been shown that  the correct Bekenstein-Hawking entropy of the near extremal five dimensional black hole is reproduced by counting the number of states in the effective string model \cite{Callan:1996dv,Horowitz:1996ay,Maldacena:1996ds}. In addition,  the Hawking  radiation of minimal scalars  has been correctly explained by   the effective string model \cite{Dhar:1996vu,Das:1996wn,Das:1996jy,Gubser:1996xe,Maldacena:1996ix,Callan:1996tv,Klebanov:1997cx}.
Although the effective string model does not correctly produce fixed scalar emissions of the black hole \cite{Krasnitz:1997gn}, it is still useful to discuss a coupling of the black hole with some minimal scalars on the string theory side. More appropriate treatment of the D1-D5-P system is given by a $\mathcal{N}=(4,4)$ superconformal field theory living on a circle \cite{David:2002wn}.

In this paper, I discuss a correspondence between string theory and the black hole membrane paradigm in the context of the D1-D5-P system. The membrane paradigm states that a distant observer sees a fictitious membrane or fluid with some  transport coefficients such as viscosities and conductivities on a stretched horizon of a black hole \cite{Thorne:1986iy,Parikh:1997ma}. Recently, we have found  that the membrane shear viscosity  of a neutral  black hole agrees with the shear viscosity of  highly excited fundamental string states at the correspondence point  if the central charge $c$ is $6$  \cite{Sasai:2010pz}. This work has been generalized and I have shown that except for the bulk viscosity, the membrane transport coefficients of an electric NS-NS 2-charged black hole correspond to the transport coefficients of the fundamental string states with a Kaluza-Klein momentum and a winding number at the correspondence point if $c=6$ \cite{Sasai:2011ys}. From these results, we can guess that in the D1-D5-P system, the membrane paradigm can be correctly explained by the effective string model because the central charge of the effective string model is 6. In fact, I show that the membrane transport coefficients of the D1-D5-P black hole induced by two kinds of  minimal scalars are  exactly the same as the corresponding transport coefficients of the  effective string model.

This paper is organized as follows. In section \ref{sec:membrane}, I review the D1-D5-P black hole and calculate the membrane transport coefficients induced by  the minimal scalars. In section \ref{sec:d1d5p}, I introduce the effective string model of the D1-D5-P system and calculate the transport coefficients induced by the minimal scalars by using the Kubo formula. Then, I find that both the transport coefficients are exactly equal.
The final section is devoted to the summary and comments.

\section{Membrane transport coefficients} \label{sec:membrane}
\subsection{D1-D5-P black hole}
Let us consider type IIB string theory compactified on $T^4\times S^1$ and wrap $n_5$ D5-branes on $T^4\times S^1$ and $n_1$ D1-branes on $S^1$. We also put  $\frac{n_p}{R}$ left-moving momentum along the D1-branes, where $R$ is the radius of $S^1$. This system becomes  a five-dimensional extremal black hole with three charges at  strong string coupling $g_s$ \cite{Callan:1996dv}.

The Einstein metric of the five-dimensional extremal black hole is given by \cite{Callan:1996dv,David:2002wn}
\begin{align}
ds^2=-f(r)^{-2/3}dt^2+f(r)^{1/3}(dr^2+r^2d\Omega_3^2), 
\end{align}
where 
\begin{align}
f(r)&=f_1(r)f_5(r)f_p(r), \\
f_{x}(r)&=1+\frac{r_{x}^2}{r^2}, ~~~~~~~~~(x=1,5,p), \\
r_{x}^2&=c_xn_x, \\
c_1=\frac{g_s\alpha'}{\vtilde},~~&~~c_5=g_s\alpha',~~~~c_p=\frac{g_s^2\alpha^{'2}}{\vtilde R^2}.
\end{align}
Here,  $V=(2\pi)^4\alpha^{'2}\vtilde$ is the volume of $T^4$. The event horizon is located at $r=0$.

To discuss the membrane paradigm, we need a finite radius of the event horizon. The generalization to the nonextremal case is given by the following Einstein metric \cite{Horowitz:1996ay,David:2002wn}:
\begin{align}
ds^2=-h(r)f(r)^{-2/3}dt^2+f(r)^{1/3}(h(r)^{-1}dr^2+r^2d\Omega_3^2), \label{eq:solution}
\end{align}
where 
\begin{align}
h(r)&=1-\frac{r_0^2}{r^2}, \\
f(r)&=f_1(r)f_5(r)f_p(r), \\
f_{x}(r)&=1+\frac{r_{x}^2}{r^2},  ~~~~~~~~~(x=1,5,p), \\
r_{x}^2&=r_0^2\sinh^2\alpha_{x},
\end{align}
and $r_0$ is the horizon radius. The mass and three charges are
\begin{align}
M&=\frac{R\vtilde r_0^2}{2\alpha^{'2}g_s^2}(\cosh 2\alpha_1+\cosh 2\alpha_5+\cosh 2\alpha_p), \\
Q_{x}&=\frac{r_0^2\sinh 2\alpha_{x}}{2c_{x}}. 
\end{align}
The extremal limit corresponds to the limit $r_0\to 0$ with at least one of $\alpha_x\to \infty$, keeping $R$, $\vtilde$ and the associated charges fixed.

This nonextremal black hole can be formally viewed as a system which is composed of noninteracting branes, antibranes and left-right moving momentum \cite{Horowitz:1996ay}.
The numbers of D1-branes, D5-branes, left-moving momentum and their anticounterparts ($\overline{\mathrm{D1}}$-branes, $\overline{\mathrm{D5}}$-branes and right-moving momentum) are
\begin{align}
n_{x}=\frac{r_0^2e^{2\alpha_{x}}}{4c_{x}},~~~~~~~\nbar_{x}=\frac{r_0^2e^{-2\alpha_{x}}}{4c_{x}}.
\end{align}
In terms of these numbers, the mass and charges are expressed by
\begin{align}
M&=\frac{R}{g_s\alpha'}(n_1+\nbar_1)+\frac{\vtilde R}{g_s\alpha'}(n_5+\nbar_5)+\frac{1}{R}(n_p+\nbar_p), \\
Q_x&=n_x-\nbar_{x}.
\end{align}

The area of the horizon is 
\begin{align}
A_H&=2\pi^2r_0^3\cosh \alpha_1\cosh \alpha_5 \cosh \alpha_p \notag \\
&=8\pi G_5(\sqrt{n_1}+\sqrt{\nbar_{1}})(\sqrt{n_5}+\sqrt{\nbar_{5}})(\sqrt{n_p}+\sqrt{\nbar_{p}}), \label{eq:area}
\end{align}
where 
\begin{align}
G_5=\frac{\pi g_s^2 \alpha^{'2}}{4\vtilde R}
\end{align}
is the five-dimensional Newton constant. Thus, the Bekenstein-Hawking entropy is \cite{Horowitz:1996ay}
\begin{align}
S_{BH}=\frac{A_H}{4G_5}=2\pi (\sqrt{n_1}+\sqrt{\nbar_{1}})(\sqrt{n_5}+\sqrt{\nbar_{5}})(\sqrt{n_p}+\sqrt{\nbar_{p}}).
\end{align}

In this paper, we assume the dilute gas regime \cite{Maldacena:1996ix},
\begin{align}
r_1, r_5 \gg r_0, r_p, \label{eq:dilute}
\end{align}
and the near extremality,
\begin{align}
n_p\gg \nbar_p. \label{eq:nearextremality}
\end{align}
The near extremality  will be necessary for perturbative string calculations to be valid at the strong coupling regime.

Then, the area of the horizon and the Bekenstein-Hawking entropy become
\begin{align}
A_H&=8\pi G_5(\sqrt{n_1n_5n_p}+\sqrt{n_1n_5\nbar_{p}}), \label{eq:area2} \\
S_{BH}&=2\pi (\sqrt{n_1n_5n_p}+\sqrt{n_1n_5\nbar_{p}}), \label{eq:entropy}
\end{align}
because $\nbar_1, \nbar_5 =0$.

\subsection{Membrane transport coefficients induced by minimal scalars}
We consider the fluctuations of the off-diagonal metric components of $T^4$ and the six-dimensional dilaton around the near extremal black hole solution, which are denoted by   $h_{ij}\equiv  f_1^{-1/2}f_5^{1/2}\delta G_{ij}~(i,j=6,7,8,9)$ and $\phi$, respectively. They are called  minimal scalars. The  action for these scalars is given by \cite{Callan:1996tv,David:2002wn}
\begin{align}
S=\frac{1}{16\pi G_5}\int d^5x \sqrt{-g}\bigg[-\frac{1}{4}\sum_{\stackrel {\scriptstyle i,j=6}{\scriptstyle i\neq j}}^9\del_{\mu}h_{ij}\del^{\mu}h_{ij}-\del_{\mu}\phi\del^{\mu}\phi\bigg],
\end{align}
where $\mu, \nu=0,1,2,3,4$.

Let us calculate the membrane transport coefficient of the near extremal black hole induced by $h_{ij}$ \cite{Parikh:1997ma, Sasai:2011ys, Iqbal:2008by}. It is enough to show it in the case of $h_{67}$. Let us set $h_{67}\equiv h$. By varying the action with respect to $h$, one finds the following boundary term on the horizon surface $\Sigma$:
\begin{align}
\delta S&=\frac{1}{16\pi G_5}\int_{\Sigma} d^4x \sqrt{-\gamma}n^{\mu} \delta h\nabla_{\mu}h,
\end{align}
where $\gamma_{\mu\nu}$ is the induced metric on  $\Sigma$. 
This boundary term is unnecessary for the bulk equation of motion to hold on $\Sigma$.
To cancel this boundary term, we add the following surface term to the action:
\begin{align}
S_{surf}=\int_{\Sigma}d^4x \sqrt{-\gamma}J_hh.
\end{align}
Then, we find
\begin{align}
J_h=-\frac{1}{16\pi G_5}n^{\mu}\nabla_{\mu}h.
\end{align}

$J_{h}$ is interpreted as a charge density on the stretched horizon induced by the bulk  field $h$. 
Since the Einstein metric of the black hole solution (\ref{eq:solution}) takes the following form,
\begin{align}
ds^2=-g_{tt}(r)dt^2+g_{rr}(r)dr^2+f^{1/3}(r)r^2d\Omega_{3}^2, \label{eq:genblackbg}
\end{align} 
the membrane charge density becomes
\begin{align}
J_{h}= -\frac{1}{16\pi G_5}\frac{1}{\sqrt{g_{rr}}}\del_r h \big|_{\Sigma}.
\end{align}
In general, fields measured by a free-falling observer must be regular at an event horizon \cite{Thorne:1986iy, Parikh:1997ma}. This is equivalent to the fact that the fields at the event horizon depend only on the ingoing null coordinate $v$ defined by \cite{Iqbal:2008by}
\begin{align}
dv=dt+\sqrt{\frac{g_{rr}}{g_{tt}}}dr.
\end{align}
Thus, near the horizon, we find
\begin{align}
\del_rh\simeq \sqrt{\frac{g_{rr}}{g_{tt}}}\del_th.
\end{align}
Therefore, the membrane charge density becomes
\begin{align}
J_{h}\simeq -\frac{1}{16\pi G_5}\frac{1}{\sqrt{g_{tt}}}\del_t h \big|_{\Sigma}=-\frac{1}{16\pi G_5}U^{\mu}\del_{\mu}h, \label{eq:chargedensity}
\end{align}
where $U^{\mu}$ is the velocity vector of an observer at the stretched horizon. 

If we assume that  $h$ is isotropic, the total  membrane charge induced by $h$ per unit time is found to be
\begin{align}
J^{tot}_h&=-\frac{A_H}{16\pi G_5}U^{\mu}\del_{\mu}h \notag \\
&=-\frac{1}{2}(\sqrt{n_1n_5n_p}+\sqrt{n_1n_5\nbar_{p}})U^{\mu}\del_{\mu}h, \label{eqtotalcharge}
\end{align}
where we have used  (\ref{eq:area2}). Therefore, the membrane transport coefficient induced by $h$ is\footnote{Since the conventional  definition of the membrane transport coefficient is given by the membrane charge density (\ref{eq:chargedensity}), the conventional membrane transport coefficient is $\frac{1}{16\pi G_5}$. However, we use   (\ref{eqtotalcharge})  to compare the membrane paradigm with the transport coefficient of the effective string model.}
\begin{align}
\Xcal^{mb}_{h}=\frac{1}{2}(\sqrt{n_1n_5n_p}+\sqrt{n_1n_5\nbar_{p}}). \label{bhh}
\end{align}
Divided by the Bekenstein-Hawking entropy (\ref{eq:entropy}), we obtain
\begin{align}
\frac{\Xcal_h^{mb}}{S_{BH}}=\frac{1}{4\pi}.
\end{align}

In the same way, the membrane transport coefficient induced by $\phi$ is
\begin{align}
\Xcal^{mb}_{\phi}&=\sqrt{n_1n_5n_p}+\sqrt{n_1n_5\nbar_{p}}, \label{bhphi} \\
\frac{\Xcal_{\phi}^{mb}}{S_{BH}}&=\frac{1}{2\pi}.
\end{align}

\section{Transport coefficients of D1-D5-P system} \label{sec:d1d5p}
\subsection{Effective string model}
The effective string model of the D1-D5-P system is described by a single D1-brane wrapped $n_1n_5$ times on $S^1$. The D1-brane has $\frac{n_p}{R}$ left-moving momentum and $\frac{\nbar_p}{R}$ right-moving momentum which are carried by the open strings attached on the D1-brane. These open strings are assumed to oscillate only inside $T^4$. This model is valid when $\vtilde\sim \mathcal{O}(1)$, $R\gg \sqrt{\alpha'}$ and the energy scale is much lower than the string  scale  \cite{David:2002wn}.

The low energy effective dynamics in our interest is given  by the following DBI action \cite{Das:1996wn, Callan:1996tv},
\begin{align}
S=-\Teff\int d^2\sigma e^{-\phi_{10}}\sqrt{-\det \gamma_{\alpha\beta}}, 
\end{align}
where $\Teff$ is the effective tension of the D1-brane, $\phi_{10}$ is the 10-dimensional dilaton and $\gamma_{\alpha\beta}~(\alpha,\beta=0,1)$ is the induced metric on the D1-brane. 

Let us choose the static gauge $\sigma^0\equiv \tau=X^0,~\sigma^1\equiv \sigma=X^5$. Expand the action around  the flat backgrounds and carrying out the Kaluza-Klein reduction of the external fields, we find \cite{Callan:1996tv}
\begin{align}
S&=S_0+S_1+\cdots, \\
S_0&=\frac{\Teff}{2}\int d^2\sigma (\Xdot^i\Xdot_i-X^{'i}X_i'), \label{eq:free} \\
S_1&=\frac{\Teff}{2}\int d^2\sigma [h_{ij}(X^{\mu})P^{ij}-\phi(X^{\mu})P^i{}_i],
\end{align}
where 
\begin{align}
\Xdot^i&=\frac{\del X^i}{\del \tau},~~~~~X^{'i}=\frac{\del X^i}{\del \sigma}, \\
P^{ij}&=\Xdot^i\Xdot^j-X^{'i}X^{'j},
\end{align}
and $S_1$ is the leading source terms of $h_{ij} ~(i\neq j)$ and $\phi$. Assuming that  the external fields $h_{ij}$ and $\phi$ depend only on time $t$ \cite{Sasai:2011ys}, $S_1$ becomes
\begin{align}
S_1&=\frac{\Teff}{2}\int dt \int_0^{2\pi Rn_1n_5} d\sigma [h_{ij}(t)P^{ij}-\phi(t)P^i{}_i]|_{\tau=t}, \notag \\
&=\int dt \bigg[\frac{1}{2}h_{ij}(t)\Jcal_h^{ij}(t)+\phi(t)\Jcal_{\phi}(t)\bigg],
\end{align}
where  
\begin{align}
\Jcal_h^{ij}(t)&=\Teff\int_0^{2\pi Rn_1n_5}d\sigma P^{ij}|_{\tau=t}, \label{eq:jij} \\
\Jcal_{\phi}(t)&=-\frac{\Teff}{2}\int_0^{2\pi Rn_1n_5}d\sigma P^{i}{}_i|_{\tau=t}. \label{eq:jphi}
\end{align}
We note that the mass dimension of $\Jcal_h^{ij}$ and $\Jcal_{\phi}$ is 1, which is the same as (\ref{eqtotalcharge}).

From the kinetic term (\ref{eq:free}), we can quantize $X^i$ in the same way as the bosonic string theory. Since $\sigma$ is identified with $\sigma+2\pi Rn_1n_5$, the mode expansion of $X^i$ becomes
\begin{align}
X^i(\tau,\sigma)=i(4\pi \Teff)^{-1/2}\sum_{m\neq 0}\bigg[\frac{\alpha_m^i}{m}e^{-i\frac{m}{Rn_1n_5}(\tau-\sigma)}+\frac{\alphat_m^i}{m}e^{-i\frac{m}{Rn_1n_5}(\tau+\sigma)}\bigg],
\end{align}
where
\begin{align}
[\alpha_m^i,\alpha_{n}^j]=[\alphat_m^i,\alphat_{n}^j]&=m\delta_{m+n,0}\delta^{ij}, \notag \\
[\alpha_m^i,\alphat_{n}^j]&=0.
\end{align}
Inserting the mode expansion into (\ref{eq:jij}) and (\ref{eq:jphi}), we find
\begin{align}
\Jcal_h^{ij}(t)&=\frac{1}{Rn_1n_5}\sum_{m\neq 0}(\alpha_m^i\alphat_m^j+\alphat_m^i\alpha_m^j)e^{-i\frac{2m}{Rn_1n_5}t}, \\
\Jcal_{\phi}(t)&=-\frac{1}{2Rn_1n_5}\sum_{m\neq 0}(\alpha_m^i\alphat_{mi}+\alphat_m^i\alpha_{mi})e^{-i\frac{2m}{Rn_1n_5}t}.
\end{align}

The mode expansion shows that each quantum which is labeled by $m$ and $i$ carries the momentum $\frac{m}{Rn_1n_5}$. Therefore, the total left-moving momentum and right-moving momentum are
\begin{align}
\frac{n_p}{R}=\frac{N_L}{Rn_1n_5},~~~~~~\frac{\nbar_p}{R}=\frac{N_R}{Rn_1n_5},
\end{align}
where $N_L$ and $N_R$ are the excitation levels of the left movers and right movers, respectively. Thus, we obtain
\begin{align}
N_L=n_1n_5n_p,~~~~~N_R=n_1n_5\nbar_p. \label{eq:valuenlnr}
\end{align}
Because of the near extremality (\ref{eq:nearextremality}), we find $N_L\gg N_R$.

The Hamiltonian of this system is
\begin{align}
H=\frac{1}{Rn_1n_5}(N_L+N_R)=\frac{n_p}{R}+\frac{\nbar_p}{R}.
\end{align}

\subsection{Transport coefficients of  effective string model}
To describe the mixed states of the effective string model, we introduce the following density matrix   \cite{Halyo:1996xe}:
\begin{align}
\rho=Z^{-1}\exp(-\beta_LN_L-\beta_RN_R),
\end{align}
where $Z=tr[\exp(-\beta_LN_L-\beta_RN_R)]$ and $\beta_{L,R}$ are the conjugate parameters of $N_{L,R}$, respectively. The mean values of the oscillation levels and the entropy are
\begin{align}
\bar{N}_{L}&\equiv \langle N_{L} \rangle =\frac{c\pi^2}{6\beta_{L}^2}, ~~~~~\bar{N}_{R}\equiv \langle N_{R} \rangle =\frac{\tilde{c}\pi^2}{6\beta_{R}^2}, \label{nlnr} \\
S&=-\langle \ln \rho \rangle =2\pi\bigg(\sqrt{\frac{c\bar{N}_L}{6}}+\sqrt{\frac{\tilde{c}\bar{N}_R}{6}}\bigg),
\end{align}
where $\langle \mathcal{O} \rangle \equiv tr (\rho \mathcal{O})$. Since there are four bosonic oscillations and four fermionic oscillations, the central charges are  $c=\tilde{c}=6$. Therefore, the entropy becomes
\begin{align}
S=2\pi(\sqrt{n_1n_5n_p}+\sqrt{n_1n_5\nbar_p}),
\end{align}
which exactly agrees with the Bekenstein-Hawking entropy (\ref{eq:entropy}) \cite{Callan:1996dv,Horowitz:1996ay}.
The statistical description is valid if $\beta_{L,R}\ll 1$ \cite{Sasai:2011ys,Halyo:1996xe}. Thus, together with the near extremality, we need $\beta_L \ll \beta_R \ll 1$ or $1 \ll \nbar_p \ll n_p$. This gives the microscopic reason of why the membrane paradigm does not exist in the extremal black hole.

Let us define the following function:
\begin{align}
f_{\Acal\Bcal}(t-t')=\frac{1}{2}\langle :[\Acal(t),\Bcal(t')]:\rangle,
\end{align}
where $\Acal(t), \Bcal(t)$ are some operators and   $:~:$ denotes the normal ordering.\footnote{The normal ordering must be evaluated after the calculation of the commutator.} A transport coefficient is obtained by
\begin{align}
\Xcal_{\Acal\Bcal}=\lim_{\omega\to 0}\frac{f_{\Acal\Bcal}(\omega)}{\omega},
\end{align}
where $f_{\Acal\Bcal}(\omega)$ is the Fourier transformation of $f_{\Acal\Bcal}(t)$. This is known as the Kubo formula \cite{Sasai:2011ys,Chaikin:book}.

Let us calculate the transport coefficient of the effective string model induced by $h\equiv h_{67}$.
Using the following formulas,
\begin{align}
\langle :\alpha_m^i\alpha_n^j:\rangle&=\frac{|n|}{e^{\beta_L|n|}-1}\delta^{ij}\delta_{m+n,0}, \\
\langle :\alphat_m^i\alphat_n^j:\rangle&=\frac{|n|}{e^{\beta_R|n|}-1}\delta^{ij}\delta_{m+n,0},
\end{align}
we find
\begin{align}
f_{\Jcal_h^{ij}\Jcal_h^{i'j'}}(t-t')&=\frac{1}{2}\langle:[\Jcal_h^{ij}(t),\Jcal_h^{i'j'}(t'):]\rangle \notag \\
&=\frac{1}{(Rn_1n_5)^2}\delta^{ij,i'j'}\sum_{m\neq 0}e^{-i\frac{2m}{Rn_1n_5}(t-t')}m\bigg(\frac{|m|}{e^{\beta_L|m|}-1}+\frac{|m|}{e^{\beta_R|m|}-1}\bigg) \notag \\
&=\frac{-2i}{(Rn_1n_5)^2}\delta^{ij,i'j'}\sum_{m=1}^{\infty}m^2\bigg(\frac{1}{e^{\beta_Lm}-1}+\frac{1}{e^{\beta_Rm}-1}\bigg)\sin \bigg(\frac{2m}{Rn_1n_5}(t-t')\bigg),
\end{align}
where $\delta^{ij,i'j'}\equiv \delta^{ii'}\delta^{jj'}+\delta^{ij'}\delta^{ji'}$. The Fourier transformation of $f_{\Jcal_h^{ij}\Jcal_h^{i'j'}}(t)$ is
\begin{align}
f_{\Jcal_h^{ij}\Jcal_h^{i'j'}}(\omega)&=\int_{-\infty}^{\infty}dt f_{\Jcal_h^{ij}\Jcal_h^{i'j'}}(t)e^{i\omega t} \notag \\
&=\frac{\pi Rn_1n_5}{4}\delta^{ij,i'j'}\omega^2\bigg(\frac{1}{e^{\beta_LRn_1n_5\omega/2}-1}+\frac{1}{e^{\beta_RRn_1n_5\omega/2}-1}\bigg).
\end{align}
Therefore, using (\ref{nlnr}) and (\ref{eq:valuenlnr}),  the transport coefficient induced by $h$ is
\begin{align}
\Xcal_h^{str}&=\lim_{\omega\to 0}\frac{f_{\Jcal_h^{67}\Jcal_h^{67}}(\omega)}{\omega} \notag \\
&=\frac{1}{2}(\sqrt{n_1n_5n_p}+\sqrt{n_1n_5\nbar_p}),
\end{align}
which exactly agrees with the membrane transport coefficient (\ref{bhh}).

In the same way, we obtain the  transport coefficient induced by $\phi$,
\begin{align}
\Xcal_{\phi}^{str}=\sqrt{n_1n_5n_p}+\sqrt{n_1n_5\nbar_p},
\end{align}
which exactly agree with (\ref{bhphi}).

\section{Summary and comments}
I have calculated the transport coefficients of the D1-D5-P system induced by two kinds of minimal scalars $h_{ij}$ and $\phi$ by using the effective string model. Then, I have found that these transport coefficients exactly agree with the corresponding membrane transport coefficients of the D1-D5-P black hole.

Two comments are in order. First, there are the other kinds of minimal scalars whose coupling with the D1-D5-P system can not be found in the effective string model \cite{David:2002wn}. Also, generically we can not use the effective string model to study the coupling with   fixed scalars \cite{Krasnitz:1997gn,David:2002wn}. Since it is known that the correct couplings of the D1-D5-P system with these scalars are given by  a $\mathcal{N}=(4,4)$ superconformal field theory   \cite{David:2002wn}, we should use the superconformal field theory to calculate the remaining transport coefficients induced by the scalar fields.

Finally, the effective string model does not possess the viscosities and conductivities  because there is no fluctuation of the effective string in  the noncompact space and therefore the effective string can not couple to the bulk metric and gauge fields. This seems to conflict with the membrane paradigm because there exists the membrane viscosities and conductivities in the D1-D5-P black hole. This discrepancy  comes from the fact that the energy scale at which  the effective string model is valid is much smaller than the string energy scale. It is known that the Hawking radiation of spin-1 and spin-2 particles are suppressed  at low energy compared to the case of the scalar particles. On the string theory side, this situation corresponds to the fact that the effective string does not couple to the bulk metric and gauge fields \cite{Mathur:1997wb}.
Thus, to discuss the viscosities and conductivities of the D1-D5-P system, we will need to study the string scale physics of the D1-D5-P system.

\section*{Acknowledgments}
I would like to thank to Esko Keski-Vakkuri  for useful  comments.  
I was supported in part by the JSPS-Academy of Finland bilateral scientist exchange program.

\end{document}